\documentstyle[epsfig, twocolumn]{esapub}

\begin{document}

\setlength{\parindent}{0pt}
\setlength{\parskip}{ 10pt plus 1pt minus 1pt}
\setlength{\hoffset}{-1.5truecm}
\setlength{\textwidth}{ 17.1truecm }
\setlength{\columnsep}{1truecm }
\setlength{\columnseprule}{0pt}
\setlength{\headheight}{12pt}
\setlength{\headsep}{20pt}
\pagestyle{esapubheadings}

\title{SPECTRO-PHOTOMETRIC CONSTRAINTS ON GALAXY EVOLUTION WITH NGST}

\author{{\bf S.~Charlot} \vspace{2mm} \\
Institut d'Astrophysique de Paris, CNRS \\
98 bis Bvd Arago, F-75014 Paris, France \\
email: charlot@iap.fr \hskip0.5truecm fax: (+33) 1 44 32 80 01}

\maketitle

\begin{abstract}

The Next Generation Space Telescope (NGST) will
gather unprecedented spectro-photometric data on galaxies 
out to the highest redshifts. It is therefore crucial to identify the
spectro-photometric diagnostics within reach of NGST, which will allow
us to best constrain the history of star formation and evolution of galaxies.
The primary parameters
to be determined are the ongoing rate of star formation and stellar mass
of galaxies at all redshifts. In this context, we briefly review the
reliability of various star formation rate and mass estimators
of galaxies in a full range of redshifts, with particular emphasis on the
relative merits of optical versus near- to mid-IR observations.
\vspace {5pt} \\


\end{abstract}

\def\simlt{\mathrel{\hbox to 0pt{\lower 3.5pt\hbox{$\mathchar"218$}\hss}
      \raise 1.5pt\hbox{$\mathchar"13C$}}}
  \def\simgt{\mathrel{\hbox to 0pt{\lower 3.5pt\hbox{$\mathchar"218$}\hss}
      \raise 1.5pt\hbox{$\mathchar"13E$}}}
\section{INTRODUCTION}
NGST will explore unknown domains in the field of galaxy evolution
(see other contributions in these proceedings). Here, we focus
on the scientific drivers for these observations from the point
of vue of models designed to interpret galaxy light (from stars and
gas) in terms of physical parameters such as the star formation rate
(hereafter SFR), mass, metallicity, age, and initial mass function
(hereafter IMF). Two main points need to be addressed. First, since
NGST will open a window on the rest-frame UV to optical light of
galaxies at $z\gg1$, one must identify, based on models of the UV
to near-IR emission of nearby objects, the best tracers to constrain
the physical properties of the most distant galaxies. Second, since NGST
will also open a new window on the less familiar near-to-mid IR emission
from nearby galaxies, it is crucial to identify whether valuable
predictions can be made in this spectral range. In particular, there
may be unexplored spectral features useful for photometric redshift
determinations, which could help to compensate for the expected limited
coverage of NGST instruments in the optical.

\section{CURRENT STATE OF AFFAIRS IN SPECTRAL MODELS}

Figure~1 shows the spectral energy distribution of a model galaxy
in the wavelength range $0.1\simlt \lambda\simlt30\,\mu$m that is
potentially relevant for NGST observations of galaxies at all 
redshifts. The model corresponds to a 0.5~Gyr old stellar population 
with constant SFR, solar metallicity, and a Salpeter IMF as computed 
using the Bruzual \& Charlot (1993, 1998) population synthesis code. 
Nebular emission was computed using G.~Ferland's (1996) CLOUDY code, 
and it includes the contribution by dust. We now address the 
reliability of the models in various spectral regions.

\begin{figure}
\psfig{file=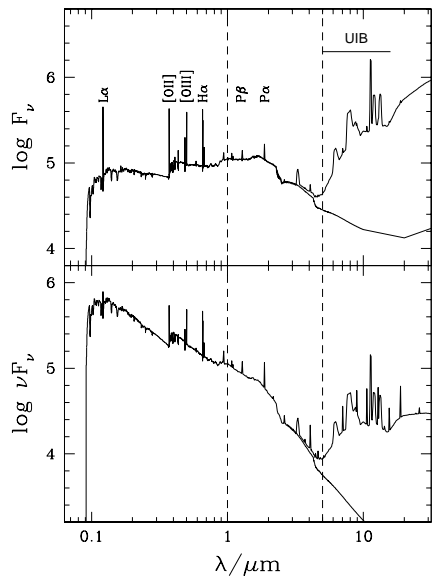,width=8cm}
\caption{\it Spectrum of a 0.5~Gyr old stellar population
with constant SFR, solar metallicity, and a Salpeter IMF as computed
using the Bruzual \& Charlot (1998) population synthesis code. The
flux is arbitrarily normalized. Nebular emission, which was computed
using Ferland's (1996) CLOUDY code, includes the contribution by
dust and the so-called unidentified infrared bands (UIB) from PAH
molecules.}
\end{figure}

Much previous work in the spectral modeling of galaxies has focused
on the $\lambda \simlt2.5\,\mu$m wavelength domain. The main reason
for this is that the stellar spectral libraries that were assembled
to build the models and the observations of galaxies the models
were aimed to interpret were mainly gathered at wavelengths $\lambda
\simlt2.5\,\mu$m (see Leitherer et al. 1996, and references therein).
A complete review of the achievements of population synthesis models in
this spectral range is beyond the scope of the present contribution (see,
e.g., Charlot 1996a for a slightly more detailed description). The most
general conclusions are that spectro-photometric fits of galaxies suffer
from several degeneracies. First, there is a degeneracy between age and
star formation timescale of models reproducing a given spectral continuum
shape, and hence, given colors. This follows from the fact that the
continuum shape is essentially driven by the ratio of blue to red stars
in a galaxy, i.e., the ratio of young to older stars. Second, there is
a degeneracy between age and metallicity in the solutions to spectral
fits of galaxies.  At fixed age, a more metal-rich stellar population
will appear redder and fainter than a more metal-poor one, while at
fixed metallicity, colors redden with age. Finally, extinction by dust
introduces an additional degeneracy in the solutions to spectro-photometric
fits of galaxies. The reason for this is that dust and metallicity (or age)
vectors are roughly parallel in many color-color diagrams.

It is worth pointing out that within the framework of a specific population
synthesis model, finer spectro-photometric diagnostics can be built 
from the $\lambda \simlt2.5\,\mu$m emission of a galaxy to constrain the
past history of star formation (in particular, by involving three-color
diagrams). In practice, however, the changes in photometric colors or
line-index strength on which such finer diagnostics 
rely are more subtle than the typical discrepancies between the predictions
of different models for the properties of stellar populations with fixed 
parameters (Charlot et al. 1996; Charlot 1996b). With current spectral
models, therefore, reconstructing the detailed history of past star 
formation from the integrated light of galaxies appears to be difficult.
An alternative and perhaps more promising approach would be to focus on the
few parameters that are generically best constrained from the integrated
light of galaxies, such as the ongoing SFR and stellar mass, and to try
and determine these parameters in galaxies at all redshifts. The important
question, then, is to identify the best tracers of these quantities to be
selected for NGST studies.

Predictions of spectral models are more uncertain in the $\lambda \simgt2\,
\mu$m wavelength domain. There are several reasons for this. First, the late
phases of stellar evolution are difficult to model because, for example, of
the uncertain influence of mass loss on the red giant branch, the difficult
treatment of convection during core-helium burning, and the occurence of
thermal pulses near the tip of the asymptotic giant branch (AGB). Moreover,
spectral (color-temperature) calibrations of theoretical stellar models
are difficult to achieve for cool stars. Empirically, determinations of 
temperatures, at least by means of lunar occultation, are limited by the
scarcity of cool stars in the solar neighborhood. Spectra of M giants
are also difficult to model theoretically because they are blanketed
mostly by molecular opacity and because their atmospheres tend to be
very extended. Since the advanced stages of stellar evolution are also
the brightest, these uncertainties limit our ability to model the
integrated spectra of galaxies in the near-IR. For example, AGB stars
are the main contributors to the integrated near-IR light of stellar
populations at ages around $0.1-1\,$Gyr.

Young stellar objects (YSOs) are other potential contributors to the
near-IR light of young galaxies. Pre-main sequence (hereafter pre-MS)
evolution of massive stars has generally been ignored in studies of
stellar populations by lack of sufficient observational constraints. 
Also, the pre-MS evolution of massive stars was once considered to be
very short, about 1\% of the MS evolution (Maeder 1996, and references
therein).  Revised models based on the accretion scenario of Palla \&
Stahler (1993, and references therein) predict that stars more massive
than $7-10\,M_\odot$ could accrete surrounding matter for much longer
than the standard contraction times, and for almost as long as the usual
MS lifetimes (Bernasconi \& Maeder 1996). Hence, young massive stars
could have burnt a substantial fraction of their central hydrogen by
the time they emerge from their parental clouds. This appears to be
supported by new statistics on the observations of young O-type stars
(see, e.g., Maeder 1996 for a review).

There is direct observational evidence that the inclusion of evolved
AGB (in particular, carbon-enriched) stars and young stellar objects
could significantly modify the predicted spectrum of Figure~1 at
wavelengths $\lambda\simgt2\,\mu$m.  Lan\c{c}on \& Wood (1997) have
performed repeated optical-infrared ($\lambda\leq2.5\,\mu$m) spectroscopy
of luminous, pulsating AGB variables in the Galaxy and LMC in order to
obtain period-averaged predictions for population synthesis models.
Preliminary results indicate that the spectral energy distributions of
these stars does not follow the pronounced decrease in $F_\nu$ predicted
around the P$\alpha$ wavelength in Figure~1. Also, $\lambda \leq2.5\,\mu$m
spectroscopy of a sample of massive YSOs by Porter et al. (1998) shows
in some cases a marked increase of the spectral energy distribution in
the same region. These observations are still too rare to be implemented
consistently in stellar population synthesis models. Among the major
difficulties to be faced, it appears that the ratio of luminous carbon
stars to oxygen-rich (bluer, M-type) stars highly depends on metallicity
(Habing 1996, and references therein). The need for more accurate
predictions in preparation for future near-to-mid infrared surveys, however,
triggers important efforts on both theoretical and observational sides.

At wavelengths redward of $5\,\mu$m the spectral energy distribution
of a galaxy can be significantly affected by dust emission (see the
contributions by D.~Elbaz and by J.-L.~Puget in these proceedings).
The amount of $\lambda\simgt5\,\mu$m emission, however, can vary 
significantly from galaxy to galaxy in no clear relation with other
properties such as morphological type or optical-IR spectral type.
This can be appreciated for example from recent observations 
with the {\it Infrared Space Observatory} ({\it ISO}) of a complete, 
optically-selected sample of 99 galaxies in the Virgo cluster by 
Boselli et al. (1998). The near- to mid-IR colors of the galaxies 
exhibit wide variations, even at similar optical to near-IR color
(see Fig.~9 of their paper). The amplitude of these variations is 
illustrated by the two possible shapes of the spectral energy 
distribution of the model galaxy spectrum in Figure~1 at
wavelengths $\lambda\simgt5\,\mu$m.

Hence, current uncertainties in the modeling of the rest-frame near-
to mid-IR spectral energy distribution of galaxies prevent us from
making reliable predictions for studies of {\it nearby} galaxies
with NGST. One of the consequences of this limitation is that strong
spectral features that could be present at wavelengths $1-5\,\mu$m 
(Figure~1) cannot yet be used to develop photometric redshift techniques
for NGST. We will return to this point in \S5. In the meantime, improving
our ability to model this spectral domain is becoming a top priority in
population synthesis studies. Dedicated observations to gather the
required stellar libraries with current infrared spectrographs is
the first necessary step towards such an improvement.

\section{SELECTING THE BEST STAR FORMATION ESTIMATORS}

The most reobust star formation estimators must be identified
to warrant the best constraints on galaxy formation and evolution to
be extracted from NGST observations. Commonly used SFR estimators are
hydrogen recombination lines (e.g., H$\alpha$), collisionally excited
oxygen lines ([OII], [OIII]), the ultraviolet continuum from
young massive stars (e.g., $L_{1500}$ at 1500~{\AA}, $L_{2800}$ at
2800~{\AA}), and the far-infrared radiation from dust heated by young
stars (e.g., Kennicutt 1992; Meurer et al. 1995; Lilly et al. 1996;
Madau et al. 1996). With NGST it will also be possible to use
near-IR hydrogen recombination lines such as P$\alpha$ at 1.88$\,\mu$m
as ordinary SFR estimators in nearby galaxies. We do not discuss here
H-Ly$\alpha$ emission as a star formation estimator
because resonant scattering of Ly$\alpha$ photons by neutral atomic
hydrogen dramatically affects their relation to the SFR in a galaxy
(see, e.g., Charlot \& Fall 1993 for a review). Also, since we are
interested in measurements of the star formation properties of galaxies
from observations at $\lambda\simlt30\,\mu$m, we do not discuss here the
far-infrared radiation by dust grains (see, e.g., Meurer et al. 1995
and references therein).

Evaluating the reliability of the above SFR estimators is difficult
because nebular properties are expected to greatly vary from galaxy
to galaxy, and even within a single galaxy. An idea of the
relative sensitivity of the different estimators to model
assumptions may be obtained by examining the range in SFR-to-light
conversion factors predicted by a wide range of models reproducing
the observed properties of nearby HII regions and galaxies. We
are in the process of completing such a study (Charlot \& Longhetti
1998), based on a combination of the Bruzual \& Charlot (1998) population
synthesis code and the latest version of Ferland's (1996) photoionization
code CLOUDY. This involves considering a range of stellar evolution
prescriptions, IMFs, star formation histories, metallicities (taken
to be the same for the stars and gas), characteric masses of ionizing star
clusters in galaxies, gas densities in HII regions, and dust contents
of HII regions. Typically, logarithmic ionization parameters are comprised
between $-3.5$ and $-1.5$, and the filling factor of the gas in HII
regions ranges from a few times 0.01 to nearly 1.

\begin{figure}
\psfig{file=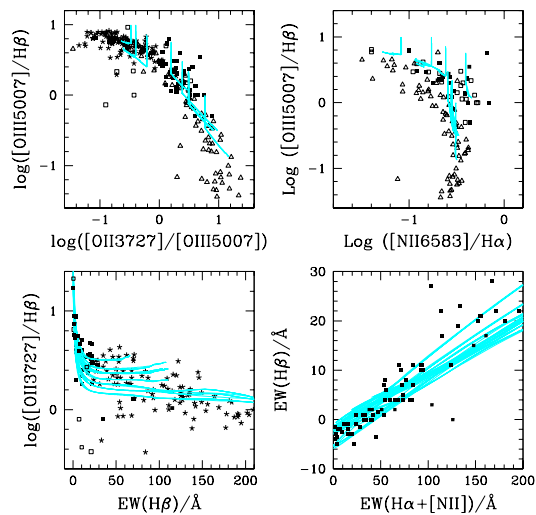,width=8cm}
\caption{\it Examples of models reproducing the emission line
properties of nearby HII regions and galaxies. The stars are
from the sample of Stasi\'nska \& Leitherer (1996), the squares from
the sample of Kennicutt (1992), and the triangles from the sample
of McCall et al. (1985). The models are described in the text.}
\end{figure}

Figure~2 shows examples of models reproducing the nebular emission
line properties of nearby HII regions and galaxies. The different
panels correspond to various combinations of flux ratios and equivalent
widths involving the H$\alpha$, H$\beta$, [OII], [OIII], and [NII]
lines. Also shown in Figure~2 are the observations of nearby HII
regions and galaxies from the samples of Kennicutt (1992), McCall et
al. (1985), and Stasi\'nska \& Leitherer (1996). Since emission line
intensities depend primarily on the amount of ionizing radiation
from short-lived, massive stars, models with different past star
formation histories and fixed other parameters lead to similar line
flux ratios in Figure~2. The predicted H$\alpha$ and H$\beta$ equivalent
widths are different, however, since these are weighted by the
continuum intensity from older stars. Figure~2 also shows that stellar
absorption at H$\beta$ dominates over nebular emission in galaxies with
declining star formation rates, leading to negative net equivalent widths
(see also Kennicutt 1992). A detailed discussion of Figure~2 and of the
dependence of the predicted nebular emission on the various model
parameters can be found in Charlot \& Longhetti (1998). Our main
purpose here is to show that the models sample reasonably well the
range of nebular properties of nearby HII regions and galaxies, and
hence, that they represent a useful basis for testing the reliability
of commonly used star formation estimators.

We now report preliminary results on the implications of this
study for SFR estimates in galaxies. We refer the reader
to Charlot \& Longhetti (1998) for a more complete
analysis. Figure~3 shows the range conversion factors $R=L/\psi$
predicted by the models described above, where $L$ stands for either
$L_{1500}$, $L_{2800}$, $L_{\rm [OII]}$, $L_{\rm [OIII]}$, $L_{{\rm
H}\alpha}$, or $L_{{\rm P}\alpha}$. We do not discuss here the
weaker, higher-order Balmer lines as SFR estimators since these can
be seriously contaminated by stellar absorption (even at H$\beta$;
see, e.g., Figure~2). The results are expressed for each estimator
in units of the reference value $R_{standard}$ corresponding to a
stellar population with constant SFR, solar metallicity, no dust and 
a Salpeter IMF truncated at 0.1 and 100$\, M_\odot$. A first 
impression from Figure~3 is that $L_{1500}$, $L_{{\rm H}\alpha}$ and
$L_{{\rm P}\alpha}$ are potentially stable star formation estimators,
with uncertainties of a factor of only 3 in $R$ despite the large
variations in model parameters.  The $L_{2800}$ luminosity, on
the other hand, stands out as a rather poor star formation estimator,
essentially for two reasons. First, it can be substantially
contaminated by the emission from older stars in galaxies with
declining star formation rates. And second, it is more sensitive
to metallicity than the $L_{1500}$ luminosity. As expected, the
collisionally-excited oxygen lines lead to larger $R$ values in
models with low metallicities because of the higher implied gas
temperatures (e.g., Kennicutt 1992). The effect is stronger for
$L_{\rm [OIII]}$ than for $L_{\rm [OII]}$.

\begin{figure}
\psfig{file=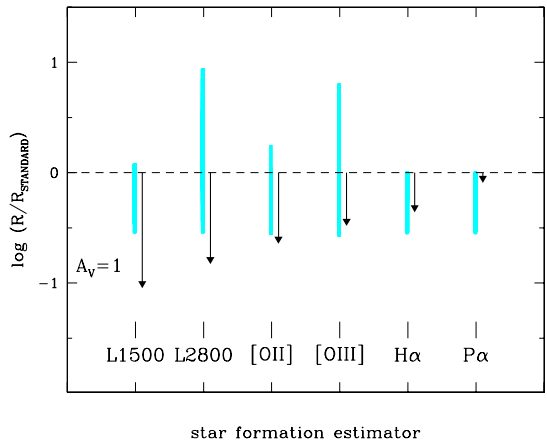,width=8cm}
\vskip-1.5truecm
\caption{\it Preliminary results on the range in light-to-SFR 
conversion factors $R=L/\psi$ from a comprehensive set of models
reproducing the observed properties of nearby HII regions and
galaxies. The results are normalized for each estimator to the
value $R_{standard}$ corresponding to a stellar population with
constant SFR, solar metallicity, no dust and a Salpeter IMF
truncated at 0.1 and 100$\, M_\odot$. Downward arrows show the
effect of extinction by a screen of Galactic-type dust with
$A_V=1$.}
\end{figure}

Extinction by dust outside the HII regions can seriously affect
the accuracy of star formation estimators in the ultraviolet.
For illustration, we show by downward arrows in Figure~3 the
effect of extinction by a screen of Galactic-type dust with
$A_V=1$. The corresponding drop in the emission at 1500~{\AA},
and hence in $L_{1500}/SFR$, amounts to a factor of 10. Also, as
emphasized by Pettini et al. (1998; see also the contribution by
Dickinson in these proceedings), uncertainties in the extinction
curves of distant galaxies have dramatic implications
for corrections of the 1500~{\AA} emission. Additional uncertainties
arise from the dependence of dust extinction on the topology of
the interstellar medium in a galaxy and on the relative mixing of
stars, gas and dust (e.g., Witt et al. 1992).

In the end, the most robust star formation estimators in Figure~3 
appear to be the H$\alpha$ and P$\alpha$ recombination lines of
hydrogen. It worth pointing out that these may also be contaminated
by nonthermal sources of radiation in galaxies, which could be 
systematically more important at high redshifts than locally
(e.g., Tresse et al. 1996). The presence of an active galactic 
nucleus in a galaxy, however, should have other readily identifiable
signatures, such as large velocity widths and strong emission lines
of highly-ionized species. Finally, we note that the uncertainties
in SFR estimates of galaxies may be reduced by constraining 
simultaneously two or more of the estimators in Figure~3. We
are investigating this effect (see Charlot \& Longhetti 1998).

\section{SELECTING THE BEST STELLAR MASS ESTIMATORS}

Dynamical mass estimates will be achievable for a subset of 
galaxies observed with NGST from either line width, rotation curve,
or galaxy-galaxy lensing measurements (see, e.g., the contributions
by Schneider and Stiavelli in these proceedings).
Stellar mass estimates, however, remain important since
they will be within reach for many more galaxies, especially at the
faintest magnitudes. Also, stellar mass relates more directly than
dynamical mass to the past history of star formation in galaxies,
and hence, it provides complementary constraints on the efficiency
of star formation as a function of dynamical environment.

The most commonly used stellar mass estimators of galaxies are the
photometric optical and near-IR magnitudes. By analogy with our
approach in the previous section, we can evaluate the relative 
accuracies of the $M/L_B$ and $M/L_K$ estimators in various redshift
ranges by considering the variations of these quantities among 
models in a wide range of metallicities and star formation histories.
We do not include nebular emission here since it negligibly affects
the integrated $B$ and $K$ magnitudes of galaxies older than a few
times $10^7\,$yr (e.g., Charlot \& Longhetti 1998). At earlier ages,
before the first red supergiants appear, the contribution to
near-IR light by hydrogen recombination continuum can significantly
affect the predicted optical to near-IR colors of stellar populations
(e.g., Leitherer \& Heckman 1995). However, this has no implication
for the results shown below because the corresponding $M/L_K$ ratios
are intermediate between those derived for stellar populations in
which the near-IR light is dominated by either red supergiants or
old, red giants.

In Figure~4 we show the ranges in rest-frame $M/L_B$ and $M/L_K$
ratios of a comprehensive set of model galaxies sampling widely
different star formation histories ($0.01 \simlt \psi_{\rm present}
/\langle \psi \rangle \simlt\,$a few) and metallicities ($0.02\simlt
Z/Z_\odot\simlt2.5$). Such models have been checked to reproduce
the observed optical to near-IR properties of galaxies of different
spectral types at various redshifts (Bruzual \& Charlot 1998, and
references therein). Since the main effect of varying the IMF
on this diagram is to change to overall normalization of the
mass-to-light ratios by changing the fraction of the total mass
in faint stars, we have adopted a standard Scalo IMF truncated
at 0.1 and 100 $M_{\odot}$. A Salpeter IMF with the same lower 
cutoff would give $M/L$ ratios a factor of $2-3$ higher. Also
indicated in Figure~4 is the separation between models younger
and older than 0.5~Gyr. Galaxies younger than 0.5~Gyr are 
expected to be rare at modest redshifts, and hence, including
these models in Figure~4 tends to overestimate the uncertainties
associated to local stellar mass estimates from photometric light. 
Extremely young galaxies, however, must be included to interpret
NGST observations at redshifts when the universe was less than 
a billion years old.

\begin{figure}
\psfig{file=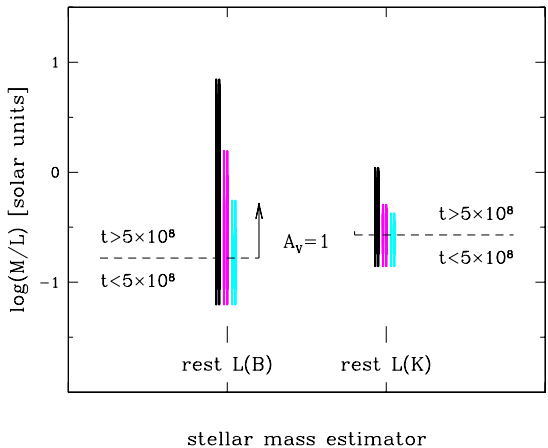,width=8cm}
\vskip-1.5truecm
\caption{\it Range in the rest-frame mass-to-light ratios
$M/L_B$ and $M/L_K$ of a comprehensive set of models reproducing
the observed optical to near-IR properties of galaxies (from 
Bruzual \& Charlot 1998). Black, heavy grey, and light grey
strips indicate the ranges in $M/L$ allowed for galaxies
no older than the age of an Einstein-de Sitter universe with
H$_0=$50$\,$km~s$^{-1}$Mpc$^{-1}$ at $z=$0, 2, and 5, respectively.
Dashed lines separate models younger and older than 0.5~Gyr,
i.e., roughly the age of the universe at $z=$8 for the adopted
cosmological model. Upward arrows (hardly distinguishable in the
case of $M/L_K$) show the effect of extinction by a screen of
Galactic-type dust with $A_V=$1.}
\end{figure}

Figure~4 inspires two main reactions. First, as expected, the
rest-frame $B$-band light is found to be a less reliable tracer
of stellar mass than the rest-frame $K$-band light, by a factor
of about 20 at $z=0$. The reason for this is that the emission
in the $B$ band is dominated by the brightest main-sequence
stars, whose characteristic luminosity decreases steadily as
a stellar population ages. In contrast, a conspiracy in the
post-main sequence evolution of stars helps to maintain the
$K$-band light at a roughly constant level in an ageing
stellar population (e.g., Charlot 1996a). Perhaps the most
instructive feature of Figure~4 is the illustration that at high
redshifts, mass estimates from the rest-frame $B$-band light are
improved with respect to $z=0$ estimates since stellar populations
in distant galaxies cannot be older than the universe. For example,
at $z=5$ the range in $M/L_B$ spanned by the models is a factor of
10 smaller than at $z=0$, while at $z\simgt8$ the allowed range
in $M/L_B$ is less than a factor of 3. The near-IR light remains
a better tracer of stellar mass than the optical light at all
redshifts, but the difference in accuracy between the two 
estimators reduces with redshift, to reach less than 50\% at
$z\simgt8$.

Therefore, although $L_K$ is confirmed to be a much better 
tracer of stellar mass than $L_B$ in galaxies at redshifts up
to a few, estimates based on the rest-frame optical light at
higher redshifts should be far more reliable than locally. 
Extinction by dust, however, will affect $B$-band emission
more significantly than $K$-band emission in galaxies (see
Figure~4). On the other hand, potential contributions to the
rest-frame near-IR emission by carbon stars (especially at the
low metallicities of distant galaxies) and young stellar objects
can affect the reliability of $M/L_K$ as a stellar mass estimator
(see \S1). It is worth mentioning again that the results of
Figure~4 were computed by adopting the local Scalo IMF for all
model galaxies. There is growing observational evidence, 
however, that the IMF might vary as a function of environment
in galaxies (e.g., Scalo 1997, and references therein). Thus,
the tight constraints on the low-mass end of the IMF that will
be obtained with NGST in nearby stellar populations (see, e.g.,
the contribution by Beckwith in these proceedings) will have
crucial implications for stellar mass estimates in the most
distant galaxies.

\section{DISCUSSION AND CONCLUSIONS}

The arguments presented in the previous sections can be used
to better assess the implications of future NGST observations
for constraining the evolution of galaxies. To visualize this
we plot in Figure~5 the model spectral energy distribution of
the star-forming galaxy computed in Figure~1 as seen at different
redshifts through NGST's $1-5\,\mu$m observational window. We
have included absorption by the Ly$\alpha$ forest of rest-frame
emission blueward of Ly$\alpha$ according to the prescription of
Madau (1995).

\begin{figure}
\psfig{file=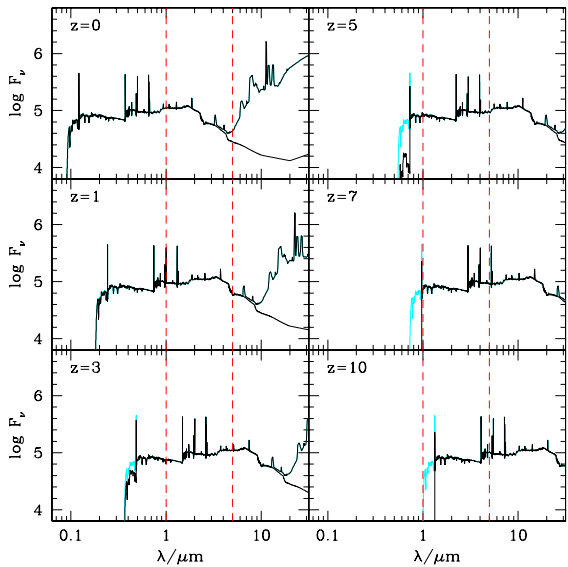,width=8cm}
\caption{\it Simple visualization of the red shift through
NGST's 1$-$5$\,\mu$m observational window of the model 
star-forming galaxy spectrum of Figure~1. The redshift is
indicated on the top left of each panel. Attenuation by the
Ly$\alpha$ forest of the emission blueward of 1216~{\AA} in
the rest frame of the galaxy has been included using the
prescription of Madau (1995). See text for discussion.}
\end{figure}

Figure~5 shows that accurate SFR estimates will be possible
with NGST for galaxies out to $z\sim7$ based on P$\alpha$ and
H$\alpha$ flux measurements. At larger redshifts, one must
appeal to higher-order Balmer lines or to the ultraviolet
emission from young stars to estimate SFRs within the 
available $1-5\,\mu$m spectral window. As noted above, however,
these alternative estimators suffer from larger uncertainties
than the P$\alpha$ and H$\alpha$ lines. Figure~5 also shows
that direct measurement of the rest-frame near-IR emission
will lead to accurate stellar mass estimates for galaxies out
to $z\sim3$. At higher redshifts the potential 5$\,\mu$m redward
boundary of the NGST observational window falls into the rest-frame
optical. As argued in \S3, this might not have a dramatic impact
on mass estimates since the accuracy of optical estimators 
is expected to increase with redshift. In particular, at
$z\sim10$, when the universe is only a few $10^8\,$yr old, mass
estimates from the rest-frame $B$-band emission should still
be fairly reliable provided that the galaxies are not heavily
obscured by dust.

Widening the observational window of NGST toward larger
wavelengths would help to improve SFR and mass estimates
of galaxies at redshifts $z\simgt5$. An extension toward
shorter wavelengths, however, presents other advantages. 
Specifically, with a short-wavelength cutoff at $1\,\mu$m
it will be difficult to single out distant galaxies from
more nearby ones using photometric redshift techniques. The
reason for this is that photometric redshift techniques
rely on the identification of strong spectral features,
such as the Lyman continuum break and the 4000~{\AA} break, 
in galaxy spectra. As described in \S1, however, at wavelengths
$1-5\,\mu$m the spectral energy distribution of galaxies is
still poorly understood in the models. Even if the particular
model presented in Figure~5 exhibits strong features in this
range of wavelengths, it would be premature to establish
photometric redshift criteria based on these features before
they are better determined. Until then, assigning photometric
redshifts to galaxies at redshifts $z \simlt1.5$, i.e., for
which the 4000~{\AA} break lies blueward of the $1-5\,\mu$m
spectral window, will be difficult. This may be of primary
concern since a galaxy like the one in Figure~5 forming stars
at a rate of $1\,M_\odot\,$yr$^{-1}$ at $z=10$ is predicted
to have an apparent magnitude $K_{AB}\approx29$, i.e.,
brighter than that of a dwarf spheroidal galaxy of total
stellar mass $10^8\,M_\odot$ at $z\approx1$. We note, however,
that so long as the {\it Hubble Space Telescope} remains
operational, it will be possible to complement NGST 
observations with space-based UV to optical observations.

Another advantage of extending NGST's observational window
further into the optical would be the local calibration of
spectral models of galaxies by means of Hertzsprung-Russell
diagram studies of nearby stellar populations. This is all
the more important in that even if SFRs and stellar masses
are determined for large samples of galaxies in a
wide range of redshifts, evolutionary links between galaxy
populations at different redshifts will not necessarily be
readily identifiable (e.g., merging history, possibility of
vanishing populations, top-heavy IMF, etc.). Therefore,
a complete understanding of the formation and evolution of
galaxies relies ultimately on our ability to relate nearby
stellar populations to the most distant galaxies. NGST
represents a milestone in the achievement of this challenge.

\section*{ACKNOWLEDGEMENTS}
I gratefully acknowledge financial support from the organizers
of this meeting. I also thank Yannick Mellier for many useful
discussions.

\end{document}